\documentclass[aps,pra,twocolumn,showpacs]{revtex4}
\usepackage{graphicx}
\usepackage{amsmath,amsfonts}

\newcommand{\er}{\mathbf{r}}
\newcommand{\ee}{{\rm e}}

\begin{document}

\title{Excited spin states and phase separation in spinor Bose-Einstein condensates}

\author{Micha\l{} Matuszewski}
\affiliation{Nonlinear Physics Center and ARC Center of Excellence
for Quantum-Atom Optics, Research School of Physical Sciences and
Engineering, Australian National University, Canberra ACT 0200,
Australia}

\author{Tristram J. Alexander}
\affiliation{Nonlinear Physics Center and ARC Center of Excellence
for Quantum-Atom Optics, Research School of Physical Sciences and
Engineering, Australian National University, Canberra ACT 0200,
Australia}

\author{Yuri S. Kivshar}
\affiliation{Nonlinear Physics Center and ARC Center of Excellence
for Quantum-Atom Optics, Research School of Physical Sciences and
Engineering, Australian National University, Canberra ACT 0200,
Australia}

\begin{abstract}
We analyze the structure of spin-1 Bose-Einstein condensates in the presence of a homogenous magnetic field. We classify the homogenous stationary states and study their existence, bifurcations, and energy spectra. 
We reveal that the phase separation can occur in the ground state of polar condensates, while the spin components of the ferromagnetic condensates are always miscible and no phase separation occurs. Our theoretical model, confirmed by numerical simulations, explains that this phenomenon takes place when the energy of the lowest homogenous state is a concave function of the magnetization. In particular, we predict that phase separation can be observed in a $^{23}$Na condensate confined in a highly elongated harmonic trap. Finally, we discuss
the phenomena of dynamical instability and spin domain formation.
\end{abstract}
\pacs{03.75.Lm, 05.45.Yv}

\maketitle

\section{Introduction}

The spin degree of freedom of spinor Bose-Einstein condensates~\cite{Stenger_Nat_1998,Ho_PRL_1998,Ohmi_JPSJ_1998} leads to a wealth of new phenomena not possessed by single-component (spin-frozen) condensates.  New spin-induced dynamics such as spin waves~\cite{Ho_PRL_1998}, spin-mixing~\cite{Chang_PRL_2004} and spin textures~\cite{Ho_PRL_1998,Leanhardt_PRL_2003} have all been predicted and observed.  These spin-dependent phenomena are possible due to the development of optical traps~\cite{StamperKurn_PRL_1998} which trap all spin components, rather than just the low-magnetic-field seeking spin states of magnetic traps.  However, the effect of an additional small non-zero magnetic field on the condensate in these optical traps was studied even in the seminal theoretical~\cite{Ohmi_JPSJ_1998} and experimental~\cite{Stenger_Nat_1998} works.  In fact the interplay of spin and magnetic field has been at the heart of some of the most impressive spinor BEC experiments, including the demonstration of spin domains~\cite{Stenger_Nat_1998}, spin oscillations~\cite{Chang_NP_2005} and spin textures and vortices~\cite{Sadler_Nat_2006}.

A spin-1 spinor BEC in a magnetic field is subjected to the well-known Zeeman effect.  At low fields the effect is dominated by the linear Zeeman effect, which leads to a Larmor precession of the spin about the magnetic field which is unaffected by spatial inhomogeneities in the condensate~\cite{Vengalattore_PRL_2007}.  At higher magnetic fields the quadratic Zeeman effect becomes important, and this can lead to much more dramatic effects in the condensate, such as coherent population exchange between spin components~\cite{Zhang_PRA_2005,Chang_NP_2005,Kronjager_PRA_2005} and the breaking of the single-mode approximation (SMA)~\cite{Zhang_NJP_2003}.  The study of the behaviour of a spin-1 condensate in the presence of a homogenous magnetic field began with the work of Stenger {\em et al.}~\cite{Stenger_Nat_1998} where the global ground state in ferromagnetic and antiferromagnetic condensates was found to be free of spin domains.  It was later found that the ground state under the constraint of fixed magnetization was significantly different to the global ground state and even broke the SMA in a harmonic trap~\cite{Zhang_NJP_2003}.
The SMA continued to be used for a homogenous condensate in a homogenous magnetic field, leading to the discovery that both phase-matched and anti-phase-matched states could exist in both ferromagnetic and antiferromagnetic condensates~\cite{Romano_PRA_2004}. Dynamical instability was found to occur in ferromagnetic condensates in nonzero magnetic field, leading to the formation of spin domains~\cite{Zhang_PRL_2005,Saito_PRA_2005} while no spin domain formation was predicted to occur~\cite{Zhang_PRL_2005} or observed~\cite{Black_PRL_2007} in antiferromagnetic condensates.  It seemed that spin domains in antiferromagnetic condensates were only to be found in the presence of inhomogenous magnetic fields~\cite{Stenger_Nat_1998}.

Recently, we have shown that, contrary to the common belief, antiferromagnetic spin-1 condensates may exhibit spin domain formation in a homogenous magnetic field~\cite{Matuszewski_PRA_2008}, provided the condensate is larger than the spin healing length.  In fact, we found that for a homogenous antiferromagnetic BEC with nonzero magnetization {\em all states are unstable}.  The form of the ground state in this case was therefore unknown.

In this work we resolve the ground state phase diagram of a spin-1 condensate in the absence of a trapping potential. We show that the translational symmetry of a homogenous BEC is spontaneously broken and phase separation occurs in magnetized polar condensates if the magnetic field is strong enough.  An analogous phenomenon has been predicted and observed previously in binary condensates \cite{Binary_PhaseSep,Timmermans}.  To explain the physics behind phase separation and determine the conditions for it to occur, we analyze the excitation spectrum of the internal spin degree of freedom of a homogenous condensate.  In contrast to spatial excitations of the condensate, which have the form of sound waves or phonons \cite{excitations, Ho_PRL_1998, Ohmi_JPSJ_1998, Ueda_SBS},
we analyze the case when the spin-dependent energy, but not the kinetic energy, is increased with respect to the ground state \cite{Romano_PRA_2004,Matuszewski_PRA_2008}.  Next, we show that for a range of experimental conditions, it is energetically favorable for the system to consist of two separate phases composed of different stationary states.  We demonstrate numerically that this phenomenon can be observed in a polar condensate trapped
in a harmonic optical potential.

The paper is organized as follows. Section~\ref{Sec_model} introduces the theoretical model of a spin-1 condensate in a homogenous magnetic field. In Sec.~\ref{Sec_SS} we study homogenous stationary states of the condensate in a magnetic field and calculate the energy of each state in terms of the magnetization and the quadratic Zeeman energy shift. We then present the internal spin excitation spectra and bifurcation behavior of the different spin states. In Sec.~\ref{Sec_gs} we analyze the ground state structure and show that phase separation can occur in polar condensates and that the behavior without a trapping potential can be used to predict the ground states in a harmonic trap.  Section~\ref{Sec_conclusions} concludes the paper.

\section{Model} \label{Sec_model}

We consider dilute spin-1 BEC in a homogenous magnetic field pointing in the $z$ direction.  The mean-field Hamiltonian of this system is given by,
\begin{align}
\label{En}
H=\int d\er \sum_{j=-,0,+}  \biggl( \frac{-\hbar^2}{2M}\nabla\psi_j^*\cdot\nabla\psi_j &+ \frac{c_0}{2} n|\psi_j|^2  \\  &+V({\bf r})|\psi_j|^2 \biggr) + H_a \nonumber
\end{align}
where $\psi_-,\psi_0,\psi_+$ are the wavefunctions of atoms in magnetic sublevels $m=-1,0,+1$, $M$ is the atomic mass, $V({\bf r})$ is an external potential and $n=\sum n_j = \sum |\psi_j|^2$ is the total atom density.  The asymmetric part of the Hamiltonian is given by,
\begin{equation}
\label{EA}
H_a = \int d\er \left(\sum_{j=-,0,+} E_jn_j + \frac{c_2}{2}|{\bf F}|^2\right)
\end{equation}
where $E_j$ is the Zeeman energy shift for state $\psi_j$ and the spin density is,
\begin{equation}
\label{spindensity}
{\bf F}=(F_x,F_y,F_z)=(\vec{\psi}^{\dagger}\hat{F}_x\vec{\psi},\vec{\psi}^{\dagger}\hat{F}_y\vec{\psi},\vec{\psi}^{\dagger}\hat{F}_z\vec{\psi})
\end{equation}
where $\hat{F}_{x,y,z}$ are the spin matrices~\cite{Isoshima_PRA_1999} and $\vec{\psi} =(\psi_+,\psi_0,\psi_-)$.  The nonlinear coefficients are given by $c_0=4\pi \hbar^2(2 a_2 + a_0)/3M$ and $c_2=4 \pi \hbar^2(a_2 -
a_0)/3M$, where $a_S$ is the s-wave scattering length for colliding atoms with total spin $S$.  The total number of atoms and the total magnetization
\begin{align}
N&=\int n d \er\,, \\
\mathcal{M}&= \int F_z d \er = \int \left(n_+ -
n_-\right) d \er\,,
\end{align}
are conserved quantities.
The Zeeman energy shift for each of the components, $E_j$ can be calculated using the
Breit-Rabi formula~\cite{Wuster}
\begin{align}
E_{\pm}& = -\frac{1}{8}E_{\rm HFS}\left(1 + 4\sqrt{1\pm \alpha + \alpha^2} \right)  \mp g_I \mu_B B\,, \nonumber \\
E_{0} &= -\frac{1}{8}E_{\rm HFS}\left(1 + 4\sqrt{1 + \alpha^2} \right)\,,
\label{BR}
\end{align}
where $E_{\rm HFS}$ is the hyperfine energy splitting at zero
magnetic field, $\alpha = (g_I + g_J) \mu_B B/E_{\rm HFS}$, where
$\mu_B$ is the Bohr magneton, $g_I$ and $g_J$ are the gyromagnetic
ratios of electron and nucleus, and $B$ is the magnetic field strength.
The linear part of the Zeeman effect gives rise to an
overall shift of the energy, and so we can remove it with the
transformation
\begin{equation}
H \rightarrow H + (N + \mathcal{M}) E_+/2 + (N - \mathcal{M}) E_-/2\,.
\end{equation}
This transformation is equivalent to the removal of the Larmor precession
of the spin vector around the $z$ axis \cite{Matuszewski_PRA_2008,Ueda_SBS}.
We thus consider only the effects of the quadratic Zeeman shift.
For sufficiently weak magnetic field we can approximate it by $\delta
E=(E_+ + E_- - 2E_0)/2  \approx \alpha^2 E_{\rm HFS}/16$, which is always
positive.

The asymmetric part of the Hamiltonian (\ref{EA}) can now be rewritten as
\begin{equation} \label{EA2}
H_{\rm a} = \int d\er \, \left(-\delta E \,n_0 + \frac{c_2}{2} |{\bf F}|^2\right)= \int d\er \, n\, e(\er)\,,
\end{equation}
where the energy per atom $e(\er)$ is given by \cite{Zhang_PRA_2005}
\begin{align} \label{EA3}
e =& -\delta E \rho_0+\frac{c_2 n}{2} |\mathbf{f}|^2 = - \delta E \rho_0 + \frac{c_2 n}{2}\left( |{\bf f}_\perp|^2 +m^2\right)\,, \nonumber\\
|{\bf f}_\perp|^2 =& 2\rho_0 (1-\rho_0) + 2\rho_0 \sqrt{(1-\rho_0)^2-m^2} \cos(\theta)\,.
\end{align}
We express the wavefunctions as $\psi_j = \sqrt{n \rho_j} \exp(i \theta_j)$ where the relative densities are $\rho_j=n_j/n$.
We also introduced the relative phase $\theta = \theta_+ + \theta_- - 2\theta_0$,
spin per atom $\mathbf{f} = \mathbf{F}/n$, and magnetization per atom $m = f_z=\rho_+-\rho_-$.  The perpendicular spin component per atom is $|{\bf f}_\perp|^2 = f_x^2+f_y^2$.

The Hamiltonian~(\ref{En}) gives rise to the Gross-Pitaevskii equations describing the mean-field dynamics of the system
\begin{align}\label{GP}
i \hbar\frac{\partial \psi_{\pm}}{\partial t}&=\left[ \mathcal{L} +
c_2 (n_{\pm} + n_0 - n_{\mp})\right] \psi_{\pm} +
c_2 \psi_0^2 \psi_{\mp}^* \,, \\\nonumber
i \hbar\frac{\partial \psi_{0}}{\partial t}&=\left[ \mathcal{L} -
\delta E + c_2 (n_{+} + n_-)\right] \psi_{0} + 2 c_2
\psi_+ \psi_- \psi_{0}^* \,,
\end{align}
where $\mathcal{L}$ is given by $\mathcal{L}=-\hbar^2\nabla^2/2M+c_0n + V({\bf r})$.

By comparing the kinetic energy with the interaction energy, we can define a characteristic
healing length $\xi=2\pi\hbar / \sqrt{2M c_0 n}$ and spin healing length
$\xi_s=2\pi\hbar / \sqrt{2M c_2 n}$. These quantities give the length scales of
spatial variations in the condensate profile induced by the spin-independent or spin-dependent
interactions, respectively. Analogously, we define magnetic healing length as
$\xi_B=2\pi\hbar / \sqrt{2M \delta E}$.

In real spinor condensates, the $a_0$ and $a_2$ scattering lengths have similar magnitude.
The spin-dependent interaction coefficient $c_2$ is therefore much smaller than its spin-independent
counterpart $c_0$. For example, this ratio is about 1:30 in a $^{23}$Na condensate and 1:220 in a $^{87}$Rb condensate far from Feshbach resonances \cite{Beata}.
 As a result, the excitations that change the total density require much more energy than those that
keep $n(\er)$ close to the ground state profile.
In our considerations we will assume that the amount of energy present in the system is not sufficient
to excite the high-energy modes, and we will treat the total atom density $n(\er)$ as a constant.

\section{Homogeneous stationary states} \label{Sec_SS}

First, we investigate the homogenous condensate in the case of a vanishing potential,
$V({\bf r}) = 0$.
We look for homogenous stationary solutions in the form
\begin{equation}
\label{stat}
\psi_{j}(\er,t) = \sqrt{n_{j}} \ee^{-i(\mu_{j} + \mu_S)t + i \theta_j}\,,
\end{equation}
where $\mu_S=c_0 n / \hbar$ is a constant.
We thus extend the studies of \cite{Romano_PRA_2004} and \cite{Matuszewski_PRA_2008}.
These solutions are stationary in the sense that the number of atoms in each magnetic sublevel is
constant in time. The relative phase between the components may change in time
as long as the phase matching condition
\begin{equation}
\mu_+ + \mu_- = 2\mu_0\,, \label{PM}
\end{equation}
is fulfilled \cite{Isoshima_PM,Matuszewski_PRA_2008}.
Because the symmetric part of the hamiltonian in~(\ref{En}) is constant, the relevant part of the hamiltonian is given by Eq.~(\ref{EA2}).

The hamiltonian~(\ref{En}) and GP equations~(\ref{GP}) are invariant under the gauge transformation $\psi_j \rightarrow \psi_j \exp(-i \beta)$
and rotation around the $z$ axis $\psi_j \rightarrow \psi_j \exp(-i F_z \gamma)$, which transform the wavefunction components according to
\begin{equation}
\label{rotations}
\left( \begin{array}{c} \psi_+ \\ \psi_0 \\ \psi_- \end{array} \right) \rightarrow
\ee^{-i\beta}\left( \begin{array}{c} \ee^{-i\gamma} \psi_+ \\ \psi_0 \\ \ee^{i\gamma} \psi_- \end{array} \right) \,.
\end{equation}
Hence the solutions can be classified using the relative densities $\rho_j=n_j/n$ and a single relative phase $\theta=\theta_+ + \theta_- - 2\theta_0$,
with the chemical potentials $\mu_j$ given as solutions to Eqs.~(\ref{GP}).  We note that for stationary solutions the relative phase must take one of two values, $\theta = 0$ or $\theta = \pi$.  We call the former ``phase-matched" states (PM) and the latter ``anti-phase-matched" (APM) states.  The names derive from the fact that within the continuum of states satisfying the spin rotations (\ref{rotations}) there is a set $(\psi_+,\psi_0,\psi_-)$ with all components in phase for the PM states, and a set with $\psi_+$ and $\psi_0$ in phase but $\pi$ out of phase with $\psi_-$ for the APM states.

The following analysis is also applicable to nonhomogeneous condensates
within the single-mode approximation (SMA), which assumes that the spin components share
the same spatial profile \cite{Ho_PRL_1998,SMA,Zhang_PRA_2005}, after replacing $n$ with $\langle n \rangle$.
This assumption is true  eg.~when the condensate size is much smaller
than the spin healing length $\xi_s$.

Stationary solutions of the system (\ref{GP}) may have one, two or three nonzero components.  We examine each case separately and then examine the existence regions and bifurcation behavior of the three states together.

\subsection{One-component solutions $(\rho_0,\,\rho_{\pm})$}

If only one component has nonzero atom density, we have two qualitatively distinct possibilities:
\begin{enumerate} \item $\rho_0=1$. This state exists for $m=0$.  From Eq.~(\ref{EA3}) and substitution of (\ref{stat}) into (\ref{GP}) we find the chemical potential of the $m=0$ component, the perpendicular spin per atom and the energy per atom,
\begin{equation}
\hbar\mu_0 = -\delta E, \quad |{\bf f}_\perp|^2=0, \quad e_{\rho_0} = -\delta E\,.
\end{equation}
\item $\rho_+=1$ or $\rho_-=1$. These two states exist for $m=1$ or $m=-1$, respectively.  Following a similar procedure to the case above we find,
\begin{equation}
\hbar \mu_+= c \quad \mathrm{or} \quad \hbar\mu_- = c, \quad |{\bf f}_\perp|^2=0, \quad e_{\rho_\pm} = \frac{c}{2}\,.
\end{equation}
where we have introduced a shorthand notation for the effective interaction coefficient, $c=c_2n$.
\end{enumerate}

\subsection{Two-component solutions $\rm (2C)$}

Here one can in general choose the vanishing component arbitrarily, but only one choice turns out to be a stationary solution.
\begin{enumerate} \item $\rho_0=0$. One distinct state exists for any values of $m$, $c$, and $\delta E$.
\begin{equation}
\hbar\mu_\pm = \pm cm, \quad |{\bf f}_\perp|^2=0, \quad e_{\rm 2C} = \frac{c}{2} m^2\,.
\end{equation}
\item $\rho_+=0$ or $\rho_-=0$. Due to the spin-dependent interaction these cases are non-stationary leading to generation of the third component, as is evident from the final terms in Eqs.~(\ref{GP}).
\end{enumerate}

\subsection{Three-component solutions $\rm (PM,\,APM)$}

We can derive the relationship between $\delta E/c$, $m$ and $\rho_0$ from the phase matching condition (\ref{PM}) and the GP equations (\ref{GP}),
\begin{equation}
\frac{\delta E}{c} = 1-2\rho_0 + s\frac{(1-\rho_0)(1-2\rho_0)-m^2}{\sqrt{(1-\rho_0)^2-m^2}} \,,
\label{crit}
\end{equation}
where $s=1$ for PM states and $s=-1$ for APM states. This condition can be alternatively derived
from the energy functional as $\partial e / \partial (\rho_0,\theta) |_{m,\delta E,n}=0$,
since the stationary states correspond to energy extrema under constraint on $m$ and $n$.
Note that $\delta E/c$ can be positive or negative depending on the sign of $c_2$.

It can be shown that if $m \neq 0$, there can be
at most one distinct PM and one APM solution from the interval $\rho_0\in [0,1]$ for a given $\delta E/c$ and $m$.  More specifically we find the following:
\begin{enumerate} \item PM solutions ($\theta=0$)
\begin{enumerate} \item For $m \neq 0$, there is one  PM solution for each $\delta E\in (-\infty, \delta E_{\rm PM})$,
where we find $\delta E_{\rm PM}=c\left(1+\sqrt{1-m^2}\right)$ from (\ref{crit}) for $\rho_0=0$.
\item For $m=0$, the solutions exist in the interval $\delta E \in (-2c,2c)$.
\end{enumerate}
\item APM solutions ($\theta=\pi$)
\begin{enumerate} \item For $m \neq 0$, there is one APM solution for each $\delta E\in (\delta E_{\rm APM},+\infty)$,
where $\delta_{\rm APM}=c\left(1-\sqrt{1-m^2}\right)$ (again found from (\ref{crit}) with $\rho_0 = 0$).
\item For $m=0$, an infinite number of degenerate solutions exist at $\delta E=0$ (no magnetic field), for any value of $\rho_0\in [0,1]$,
however APM states do not exist at nonzero field.
\end{enumerate}
\end{enumerate}
We see from (\ref{EA3}) that for three-component solutions $f_{\perp}$ is nonzero for both PM and APM states, hence these solutions
break the $U(1)$ rotational symmetry. In fact, the state investigated in \cite{Ueda_SBS} is identical to the PM
state, which is the ground state of a ferromagnetic condensate whenever $m \neq 0$ or $\delta E < 2|c|$.
Moreover, as the chemical potentials of the three components are in general not equal the spin vector
rotates around the $z$ axis, in addition to the Larmor rotation,
at a rate proportional to the difference in chemical potentials.

To obtain further information about the nature of the three-component solutions we examine some of the limiting behavior.  By expanding $\rho_0$ about $1-|m|$ we can determine the behavior in the limit $|\delta E/c| \rightarrow +\infty$,
\begin{align}
\rho_0 &= (1-|m|)- \frac{c^2}{2\delta E^2} |m| (1-|m|)^2 \,, \nonumber\\
f_{\perp}^2 &= 2|m|(1-|m|) \,, \\
e_{\rm (A)PM} &= -\delta E (1-|m|) + c |m|\left(1-\frac{|m|}{2}\right) \,,\nonumber\\
\hbar\mu_0&= - \delta E + O(c)\,, \hbar\mu_\pm = -\delta E [1  \mp {\rm sgn}(m)] + O(c)\,. \nonumber
\end{align}
On the other hand, when $\delta E \rightarrow \delta E_{\rm (A)PM}$ (close to the bifurcation point),
\begin{align}
\rho_0 &= \frac{1}{2} \frac{|\delta E - \delta E_{\rm (A)PM}|}{\delta E_{\rm (A)PM}} \sqrt{1-m^2} \,,\\
f_{\perp}^2 &\sim \rho_0, \quad e_{\rm (A)PM} = \frac{c}{2} m^2\,, \nonumber\\
\hbar \mu_0&=0\,, \quad \hbar\mu_\pm = \pm cm \,.\nonumber
\end{align}
\begin{figure}[tbp]
\includegraphics[width=8cm]{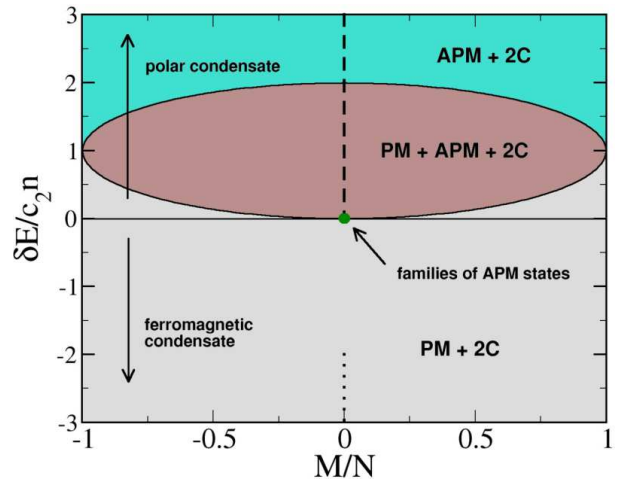}
\caption{Diagram of existence of two- (2C) and three-component (PM or APM) homogenous stationary states in
spin-1 condensates. In addition to the solutions shown on the diagram, one-component solutions
$\rho_j=1$ exist with $j=-,0,+$. The dotted and dashed lines at $\mathcal{M}=0$ indicate the
absence of a PM or APM state respectively.}\label{existence_diag}
\end{figure}

\subsection{Existence and order parameter spaces}

Combining the above results, the complete diagram of existence of the two-component and three-component states is shown in Fig.~\ref{existence_diag}.
Because $\delta E >0$, the upper half of the diagram corresponds to polar condensates ($c_2 > 0$), and the
lower half describes  ferromagnetic BECs ($c_2 < 0$).
There is clearly a region of coexistence of PM and APM solutions in polar condensates. For ferromagnetic condensates APM states only exist at zero magnetic field (and zero magnetization).

The order parameter space of the two- and three-component solutions in nonzero magnetic field is $U(1)\times U(1)$ (toroid),
due to the two symmetries, gauge and rotation around $z$ axis, which leave the phase $\theta$ and the atom density
in each of the components unchanged. However, the one-component solutions have only the $U(1)$ parameter space
(the same as for a scalar condensate), due
to the equivalence of gauge transformation and rotation. When no magnetic field is present, the
situation is different because PM and APM states become degenerate for all values of $\rho_0$ and
form the families of polar and ferromagnetic states together with $\rho_\pm$ and $\rho_0$ states, respectively (see Fig.~\ref{states}).
The order parameter manifolds of these families are $SO(3)$ for the ferromagnetic \cite{Ho_PRL_1998} and $U(1) \times S^2 / \mathbb{Z}_2$
for the polar states family \cite{Zhou,Mukerjee}.

\subsection{Internal spin excitations spectra and bifurcations} \label{Sec_spectra}

In scalar condensates, where only one spin component is present, the Bogolubov theory can be used to describe spatial excitations of
the condensate, which have the form of sound waves or phonons \cite{excitations}. In spinor condensates,
modes of a similar nature have been studied \cite{Ho_PRL_1998, Ohmi_JPSJ_1998, Ueda_SBS}, but another degree
of freedom is also available. One can consider internal spin excitations, where the spin-dependent energy, but not the kinetic energy,
is increased with respect to the ground state. In contrast to the phonon-type excitations, the excitation spectrum of the internal spin states
is discrete. The stationary solutions, described in the previous section, form a set
of such modes when a magnetic field is applied to a homogeneous condensate.
We argue that the energy of the system can be exchanged between the spin modes and spatial excitations, which has
important consequences for the condensate dynamics.

\begin{figure}
\includegraphics[width=8cm]{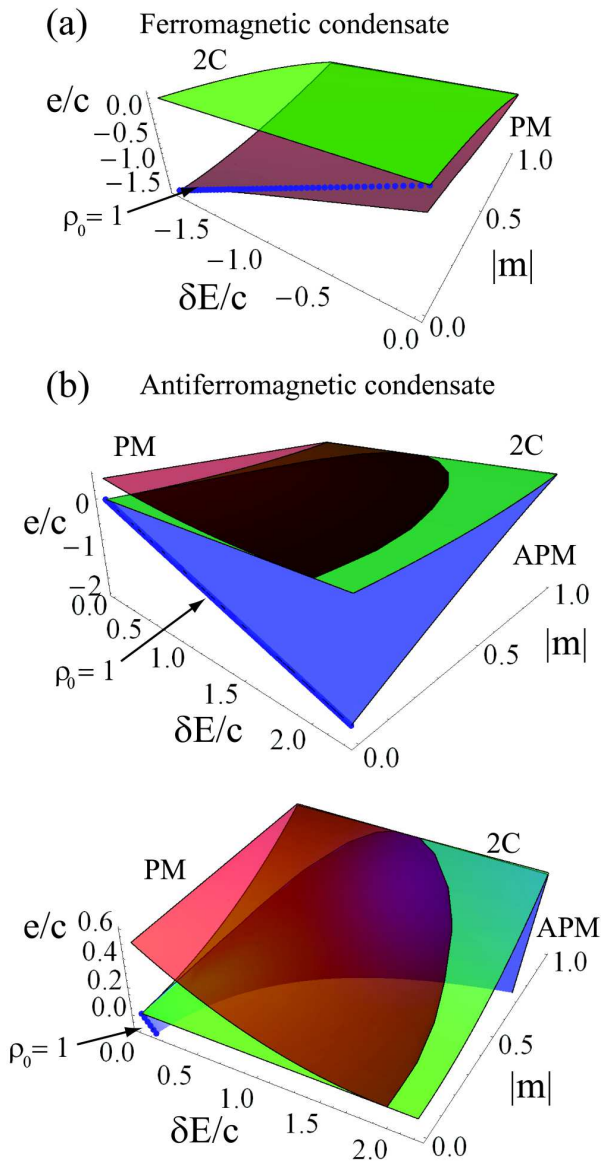}
\caption{Normalized energy per atom $e/(|c_2| n)$ as a function of magnetization $|m|=|\mathcal{M}|/N$ and normalized quadratic Zeeman
energy $\delta E/(c_2 n)$ for homogenous stationary states of spin-1 condensates, including two-component (2C), phase-matched (PM) and anti-phase-matched (APM) states.  The single-component $\rho_0 = 1$ state is also shown. (a) Ferromagnetic condensate; (b) Anti-ferromagnetic condensate, with lower panel showing a zoom of the upper panel.} \label{En_3D}
\end{figure}

\begin{figure}[tbp]
\includegraphics[width=8.5cm]{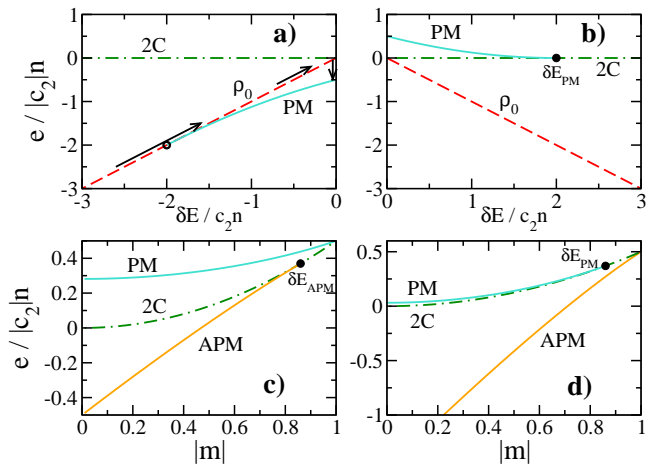}
\caption{(a,b) Normalized energy per atom $e/(|c_2| n)$ in function of normalized quadratic Zeeman
energy $\delta E/(c_2 n)$ for homogenous stationary states in the $\mathcal{M}=0$ case for (a) ferromagnetic and (b) polar condensate.
The arrows in (a) show the scenario of the quenched condensate experiment \cite{Sadler_Nat_2006}.
(c,d) Energy in function of the absolute value of the magnetization $|m|$ for (c) $\delta E/(c_2 n)=0.5$ and
(d) $\delta E/(c_2 n)=1.5$. The black dots indicate bifurcation points where $\delta E = \delta E_{\rm APM}$
and $\delta E = \delta E_{\rm PM}$, respectively. The energy of the APM state is a slightly concave function of the magnetization,
which gives rise to the phase separation (see Sec.~\ref{Sec_gs}).
} \label{En_2D}
\end{figure}

The dependence of the energy per atom given by Eq.~(\ref{EA3}) on magnetic field and magnetization for the spin states
studied in the previous section is shown on three-dimensional plots
in Fig.~\ref{En_3D}. Note that renormalized variables $e/(|c_2| n)$ and $\delta E/(c_2 n)$ are used, which allows
us to include all possible configurations of spin-1 condensate just in two universal graphs (there is no fixed parameter).

The energy dependence cross-section in the particular case $m=0$ is shown in Fig.~\ref{En_2D}(a,b). It is clear from Fig.~\ref{En_2D}(a)
that when the Zeeman energy is decreased, the $\rho_0$ state ceases to be the lowest-energy state of the ferromagnetic
condensate, and the PM state becomes the ground state~\cite{Ueda_SBS}. This fact has been utilized in the experiment by
Sadler et.~al.~\cite{Sadler_Nat_2006}, where the ferromagnetic BEC prepared in the $\rho_0$ state in strong magnetic field
was suddenly quenched to the low magnetic field regime. As the condensate relaxed locally to the PM state
in a conservative process, an amount of energy was transformed from the spin to the kinetic energy,
which allowed the formation of spin domains and topological excitations. As we will show later, this 
excess energy was necessary since no spin domains exist in the ground state of a ferromagnetic condensate.
In Fig.~\ref{En_2D}(c,d) we show dependence of energy on magnetization for fixed magnetic field strength
and indicate the points of bifurcation.

We summarise the degeneracies and bifurcations between the various spin states in the following list:
\begin{enumerate}
\item The three-component (PM, APM) and two-component (2C) solutions become identical to one-component solutions $\rho_\pm$ at $m=\pm 1$.
\item The PM states bifurcate from two-component solutions (2C) at $\delta E = \delta E_{\rm PM}$.
The APM states bifurcate from 2C solutions at $\delta E = \delta E_{\rm APM}$.
\item The APM states become identical to $\rho_0$ states at $m=0$.
\item In a ferromagnetic condensate ($c < 0$),
PM states become identical to the $\rho_0$ state at $m=0$ if $\delta E > 2|c|$. However, for $\delta E \in (0,2|c|)$ there exists a separate
PM state with $m=0$ which is not equivalent to $\rho_0$, see Fig.~\ref{En_2D}(a). Hence $\delta E = -2c$, $m=0$ is a bifurcation point.
\item At $\delta E = 0$ (no magnetic field), all the APM states have $|{\bf f}|=m=0$ and are degenerate and continuously connected to the $\rho_0$ state.
These states together form the family of polar states, which is the ground state of polar condensates \cite{Ho_PRL_1998,Zhou}.
\item At $\delta E = 0$, all the PM states (with different values of $m$) have $|{\bf f}|=1$ and become degenerate and continuously connected to the $\rho_\pm$ states.  These states form the family of ferromagnetic states, which are the ground states of ferromagnetic condensates \cite{Ho_PRL_1998}.
\end{enumerate}

\begin{figure}[tbp]
\includegraphics[width=8.5cm]{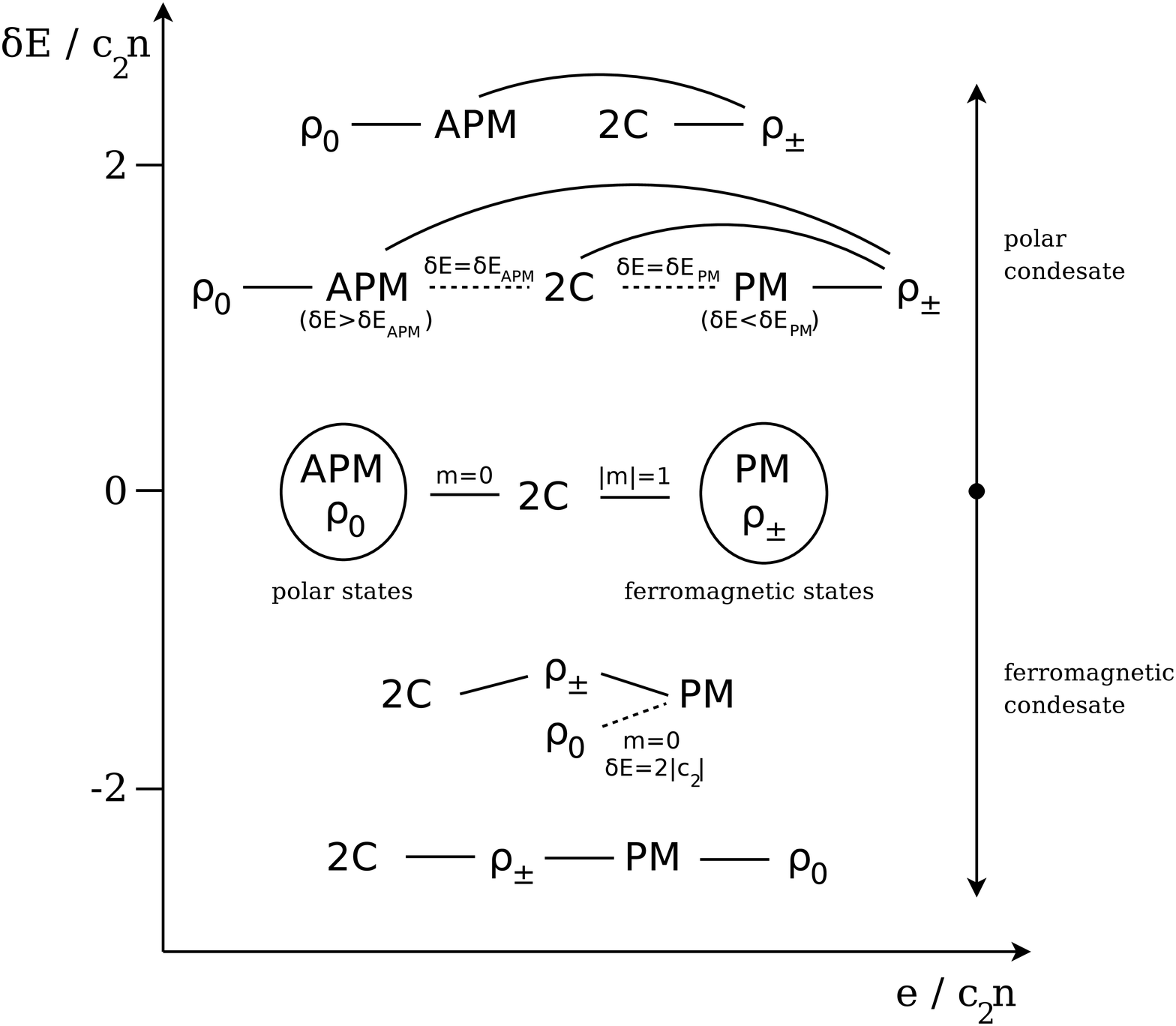}
\caption{Schematic picture of equivalences and bifurcations of homogenous stationary states in various regimes of
$\delta E/(c_2 n)$. Solid lines indicate that two states become identical for a particular value of $|m|$, and dashed lines
correspond to bifurcations occuring at a given value of $\delta E$. In the $\delta E =0$ case (no magnetic field), the two circles
show that degenerate states form the families of polar and ferromagnetic states, which are the ground states of polar
and ferromagnetic condensates, respectively \cite{Ho_PRL_1998}. Note that for ferromagnetic condensates($c_2<0$), the energy grows to the
left.} \label{states}
\end{figure}

These connections between states, together with the energy hierarchy, are schematically collected in Fig.~\ref{states}. Equivalence between two states
at $m=0$ or $|m|=1$ is indicated with continuous lines, while bifurcations (occurring with either changing $m$ or $\delta E/(c_2 n)$) are marked by dashed lines.  For example,
2C states become equivalent to $\rho_{\pm}$ states (i.e. either $\rho_+$ or $\rho_-$) whenever $m\rightarrow\pm 1$. On the other hand, APM states bifurcate from 2C states when the
quadratic Zeeman energy crosses the value $\delta E_{\rm APM}$, and separate APM and 2C states exist for the same value of $m$ for $\delta E > \delta E_{\rm APM}$.
The two circles in the middle correspond to the polar and ferromagnetic state families at zero magnetic field.

\section{Ground states and phase separation} \label{Sec_gs}

\subsection{No trapping potential}

The ground states of spin-1 condensates in homogenous magnetic field
have been studied in a number of previous works \cite{Zhang_NJP_2003,Stenger_Nat_1998,Ueda_SBS,SMA,Zhou}.
The most common procedure \cite{Stenger_Nat_1998,Ueda_SBS} involves minimization of the energy functional with constraints on the number
of atoms $N$ and the total magnetization $\mathcal{M}$. The resulting Lagrange multipliers $p$ and $q$ serve
as parameters related to the quadratic Zeeman shift $\delta E$ and the magnetization $m$. An alternative
method, elaborated in \cite{Zhang_NJP_2003}, consists of minimization of the energy functional in the parameter space
of physically relevant variables $B$ and $m$. Most of the previous studies, however, were assuming that the
condensate remains homogenous and well described by the single-mode approximation;
in particular, the spatial structure observed in \cite{Stenger_Nat_1998} resulted from the applied
magnetic field gradient, but the BEC was assumed to be well described by the homogenous model at each point in space.
In Ref.~\cite{Zhang_NJP_2003}, the breakdown of the single-mode approximation was shown numerically for a condensate confined
in a harmonic potential.

We correct the previous studies by showing that when the condensate size is larger than the spin healing length $\xi_s$,
the translational symmetry is spontaneously
broken and phase separation occurs in magnetized polar condensates if the magnetic field is strong enough.
This phenomenon takes place
when the energy of the spin state with the lowest energy is a concave function of $m$ for a given $\delta E$.
On the contrary, the energy is always a convex function of $m$ for the ferromagnetic condensate, and no phase separation occurs.
Note that phase separation has been previously predicted in binary condensates \cite{Binary_PhaseSep,Timmermans} and
in ferromagnetic condensates at finite temperature~\cite{Machida_Transitions}.

\begin{figure}
\includegraphics[width=8.5cm]{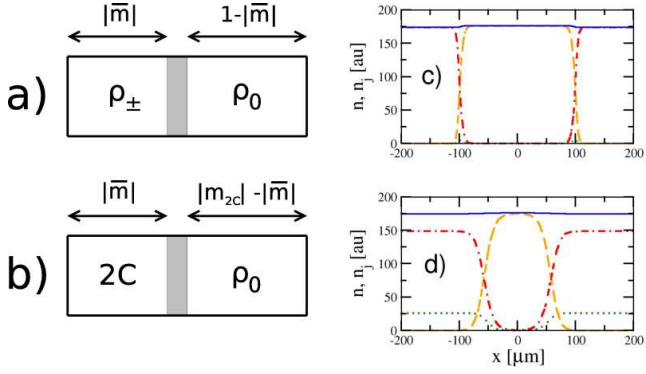}
\caption{Schematic structure of the phase separated states a) $\rho_\pm + \rho_0$ an b) ${\rm 2C} + \rho_0$.
The shaded region, in which all three components are nonzero, has the
approximate extent of one spin healing length $\xi_s$ or magnetic healing length $\xi_B$, whichever is greater.
The relative size of the domains is indicated with arrows.
(c, d) The corresponding wavefunction profiles obtained numerically with periodic boundary conditions 
in the case of $^{23}$Na for $\overline{m}=0.5$ with
(c) $\delta E / (c_2 n) =0.8$ and (d) $\delta E / (c_2 n) =0.23$.
The $n_+$,  $n_0$, and  $n_-$ components are depicted
by dash-dotted, dashed, and dotted lines, respectively. The solid lines show the total density.
} \label{domains}
\end{figure}

Two types of domain structures, depicted in Fig.~\ref{domains}, are composed of two different stationary states connected
with a shaded region where all three components are nonzero to ensure proper matching of the chemical potentials (\ref{PM}).
These two domain states have the advantage that the perpendicular spin is nonzero only in the transitory region, hence their energy is relatively
low in polar condensates. In fact, these are the only phase separated states that can be the ground states of a homogenous condensate.
Their energies per atom in the limit of infinite condensate size, which allows for neglecting of the relatively small intermediate region are
\begin{align} \label{GS_energies}
e_{\rho_\pm + \rho_0} &= |\overline{m}| e_{\rho_\pm} + (1-|\overline{m}|)e_{\rho_0} 
\,, \nonumber\\
e_{\rm 2C + \rho_0} &= \frac{\overline{m}}{m_{\rm 2C}} e_{\rm 2C}\big|_{m=m_{\rm 2C}} +\left(1-\frac{\overline{m}}{m_{\rm 2C}}\right) e_{\rho_0} \,,
\end{align}
where $\overline{m}=\mathcal{M}/N$ is the average magnetization and the magnetization of the 2C component $m_{\rm 2C}$
is a free parameter that has to be optimized to obtain the lowest energy state.

\begin{table} [tbp]
\begin{tabular}{|l|c|c|}
\hline
Condensate & Parameter range & Ground state  \\
\hline
Ferromagnetic & $ 2 \leq \frac{\delta E}{|c_2| n}$ and  $\overline{m}=0$ & $\rho_0$\\
 & $\frac{\delta E}{|c_2| n} < 2$ or  $\overline{m} \neq 0$ & $\rm PM$\\
\hline
Polar & $\overline{m}=0$ & $\rho_0$\\
 & $\frac{\delta E}{c_2 n} \leq \frac{\overline{m}^2}{2}$  & $\rm 2C$\\
 & $ \frac{\overline{m}^2}{2} < \frac{\delta E}{c_2 n} <  \frac{1}{2}$ and $\overline{m}\neq 0$ & $\rm 2C + \rho_0$\\
 & $\frac{1}{2} \leq \frac{\delta E}{c_2 n}$ and $\overline{m}\neq 0$ & $\rho_\pm + \rho_0$\\
\hline
\end{tabular}
\caption{Ground states of spin-1 condensates in homogenous magnetic field. The states $\rm 2C + \rho_0$ and $\rho_\pm + \rho_0$
correspond to phase separation (see Fig.~\ref{domains}).}\label{TGS}
\end{table}

The ground states can be determined by comparing energies of the phase separated states with the energies of the homogenous solutions of Sec.~\ref{Sec_SS}.
The results for both polar and ferromagnetic condensates are collected in Table~\ref{TGS}. In the cases when no phase separation occurs,
our results are in agreement with those obtained in~\cite{Zhang_NJP_2003}. Note that we assumed that the condensate size is much larger than $\xi_s$ and $\xi_B$. 
For small condensates, the results of~\cite{Zhang_NJP_2003} are correct.

In the case of high magnetic field strength, one of the Zeeman sublevels is practically depleted~\cite{Zhang_NJP_2003} and the condensate becomes
effectively two-component. The existence of the $\rho_\pm + \rho_0$ phase in a polar condensate can then be understood
within the binary condensate model~\cite{Binary_PhaseSep,Timmermans}. We note that the experiment reported in Ref.~\cite{Ketterle_Metastable},
performed in this regime, can be viewed as the first confirmation of phase separation in spin-1 BEC in a homogenous magnetic field. However,
the ground state was not achieved, and a multiple domain structure was observed.

\begin{figure}[tbp]
\includegraphics[width=8cm]{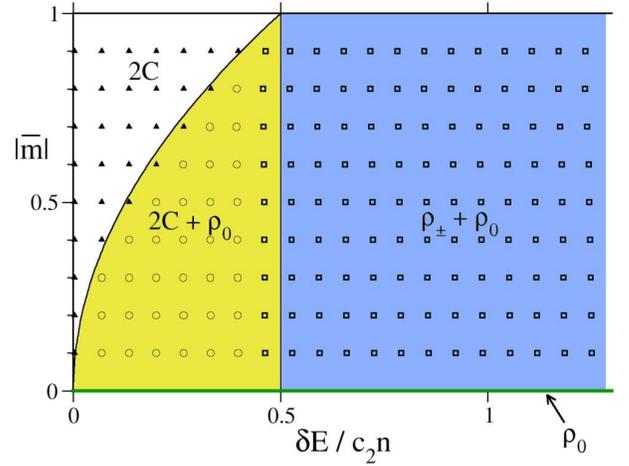}
\caption{Ground state phase diagram of the polar condensate. The symbols correspond to numerical data obtained for the parameters of $^{23}$Na, with solid triangles representing 2C, open circles 2C+$\rho_0$ and open squares $\rho_\pm + \rho_0$. The solid lines and shading are given by the analytical formulas from Table~\ref{TGS}.  } \label{phase_diagram}
\end{figure}

In Fig.~\ref{phase_diagram} we present the phase diagram of polar condensates, obtained both numerically and using analytical formulas from Table~\ref{TGS}.
The ground state profiles for a quasi-1D condensate were found numerically by solving the 1D version of Eqs.~(\ref{GP}) \cite{Matuszewski_PRA_2008}
\begin{align}\label{GP_1D}
i \hbar\frac{\partial \tilde{\psi}_{\pm}}{\partial t}&=\left[ \tilde{\mathcal{L}} +
\tilde{c}_2 (\tilde{n}_{\pm} + \tilde{n}_0 - \tilde{n}_{\mp})\right] \tilde{\psi}_{\pm} +
\tilde{c}_2 \tilde{\psi}_0^2 \tilde{\psi}_{\mp}^* \,, \\\nonumber
i \hbar\frac{\partial \tilde{\psi}_{0}}{\partial t}&=\left[ \tilde{\mathcal{L}} -
\delta E + \tilde{c}_2 (\tilde{n}_{+} + \tilde{n}_-)\right] \tilde{\psi}_{0} + 2 \tilde{c}_2
\tilde{\psi}_+ \tilde{\psi}_- \tilde{\psi}_{0}^* \,,
\end{align}
with $\tilde{\mathcal{L}} = -(\hbar^2/2m)\partial^{2}/\partial x^2 + \tilde{c}_0$, where
$\tilde{c}_0=4 \hbar\omega_{\perp}(2 a_2 + a_0)/3$, $\tilde{c}_2=4
\hbar\omega_{\perp}(a_2 - a_0)/3$, $\int dx \sum |\tilde{\psi}_j| = N$, and $\omega_\perp$ is the transverse trapping frequency.
We imposed periodic boundary conditions on $\tilde{\psi}_j(x)$ and used the parameters corresponding to
a $^{23}$Na BEC containing $N=5.2 \times 10^4$ atoms confined in a transverse trap with frequency $\omega_\perp=2\pi \times 10^3$.
The Fermi radius of the transverse trapping potential is
smaller than the spin healing length,
and the nonlinear energy scale is much smaller than the transverse
trap energy scale, which allows us to reduce the
problem to one spatial dimension \cite{Beata,NPSE}.
The solutions were found numerically using the normalized gradient flow method \cite{BaoLim},
which is able to find a state which minimizes the total energy for given $N$ and $\mathcal{M}$, and
fulfills the phase matching condition (\ref{PM}).
The stability of the resulting states was verified using numerical time evolution according to Eqs.~(\ref{GP_1D}).
The slight discrepancy between numerical and analytical results can be accounted for by the finite size of
the condensate (the box size was $\sim10 \,\xi_s$), and by the deviation from the assumption that the total density is constant (see the discussion at the end of Sec.~\ref{Sec_model}).  Due to the finite value of the ratio $c_2/c_0$ there is a slight density modulation, as is evident in Fig.~\ref{domains}(c,d).

\subsection{Condensate trapped in a harmonic optical potential}

The results from the preceding subsection can be verified experimentally in configurations
involving toroidal or square-shaped optical traps~\cite{Toroidal}. However, in most experiments on BECs, harmonic potentials are used.
The relevance of these results is not obvious in the case of harmonic trapping, since the coefficient $\delta E /(c_2 n)$,
one of the main parameters controlling the condensate properties, varies in space due to the varying total density $n$.

\begin{figure}
\includegraphics[width=8.5cm]{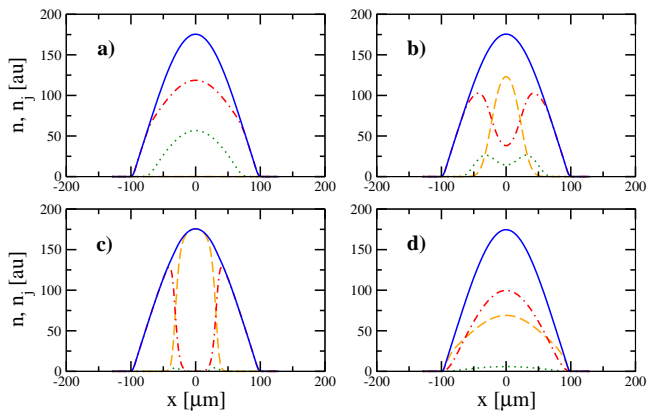}
\caption{Ground state profiles in a harmonic trap potential. Phase separation occurs
in the polar $^{23}$Na condensate when the magnetic field strength is increased from (a) $B=0.1$G, $\delta E / (c_2 n_{\rm max})=0.09$ to
(b) $B=0.12$G, $\delta E / (c_2 n_{\rm max})=0.13$ and (c) $B=0.25$G, $\delta E / (c_2 n_{\rm max})=0.56$.
For comparison, the ground state of a $^{87}$Rb condensate is
shown in (d) for $B=0.2$G, $\delta E / (c_2 n_{\rm max})=-0.41$. The $n_+$,  $n_0$, and  $n_-$ components are depicted
by dash-dotted, dashed, and dotted lines, respectively. The solid lines shows the total density.
Other parameters are $N=2.1 \times 10^3$, $\omega_\parallel=2\pi\times70$ ($^{23}$Na), $\omega_\parallel=2\pi\times48$ ($^{87}$Rb),
$\omega_\perp=2\pi\times10^3$ and $\overline{m}=0.5$.}
\label{profiles}
\end{figure}

The ground states in a highly elongated harmonic trap, where the parallel part of the potential has the form $V(x) = \frac{1}{2}M\omega_\parallel^2 x^2$,
are presented in Fig.~\ref{profiles}. We can see that as the magnetic field strength is increased,
phase separation occurs and the $\rho_\pm+\rho_0$ domain state is formed. However, in contrast to the previous case, the transition is not sharp,
and in particular there is no distinct $2C+\rho_0$ phase for any value of the magnetic field.
Note that the state in Fig.~\ref{profiles}(a) is also spatially separated due to different Thomas-Fermi radii of the $\psi_-$ and  $\psi_+$ components;
however, this is an example of potential separation, as opposed to phase separation~\cite{Timmermans}, since it is does not occur in the absence of the potential.
On the other hand, Fig.~\ref{profiles}(d) shows that the components of ferromagnetic condensate are miscible even in the regime of strong magnetic field.
In the regions where the wavefunctions overlap, the relative phase is equal to $\theta=0$ for ferromagnetic
and $\theta=\pi$ for polar ground states, since these configurations minimize the spin energy~(\ref{EA3}).

The characteristic feature of phase separation in the polar BEC is that the $m=0$ domain tends to be localized in the center of the trap,
as shown in Fig.~\ref{domains}(b) and (c).
This can be explained by calculating the total asymmetric energy of the condensate (\ref{EA2}),
again assuming that the contribution from the intermediate region connecting the domains is negligible,
\begin{align}
H_a &\approx   \int_{\rho_0} d \er \, n \left(-\delta E\right) +  \int_{\rho_\pm} d \er \, n \frac{c_2n}{2} \\\nonumber
&= - \delta E (N-|\mathcal{M}|) + \frac{c_2 N}{2} \langle n \rangle_{\rho_\pm} \,,
\end{align}
where $\langle n \rangle_{\rho_\pm}$ is the mean condensate density within the area of the $\rho_\pm$ domain. We see that the energy will be the lowest
if this domain is localized in the outer regions, where the condensate density is low.


\subsection{Spin domain formation}

Our results presented above show that the domain structure forming in polar condensates is absent in ferromagnetic BECs. This may seem to contradict the common understanding of ferromagnetism and the results of the quenched BEC experiment Ref.~\cite{Sadler_Nat_2006}. The conventional picture of a ferromagnet involves many domains pointing in various directions separated by domain walls. Similar structure has been observed in Ref.~\cite{Sadler_Nat_2006}. However, these cases correspond to the situations when there is an excess kinetic energy
present in the system, due to finite temperature or excitation of the spatial modes.
On the other hand, our study is limited to the ground states at $T=0$. It is easy to see from Eq.~(\ref{EA3}) that
in zero magnetic field the ground state of a ferromagnetic BEC will always consist of a single domain with maximum possible value of the spin vector $|{\bf f}|=1$, pointing in the same direction at all points in space. However, when the temperature is finite, more domains can be formed each with a different direction of the spin vector.

We emphasize that the domain structure of the ground state in polar condensates is very different from the domains
formed when the kinetic energy is injected in the system as in Ref.~\cite{Sadler_Nat_2006}. The latter constantly
appear and disappear in a random sequence~\cite{Sadler_Nat_2006,MurPetit,Saito_PRA_2005,Ueda_Defect,Zhang_PRL_2005,Matuszewski_PRA_2008}. On the contrary, the ground state domains are stationary and are positioned in the center of the trap. They exist in the lowest-energy state, while the dynamical domains require an amount of kinetic energy to be formed.
The ground state domains can be prepared in an adiabatic process, involving adiabatic rf sweep or a slow change of the
magnetic field~\cite{Ueda_Defect,Ketterle_Metastable}, while the kinetic domains require a sudden quench~\cite{Sadler_Nat_2006,Ueda_Defect}.

\subsection{Dynamical stability}

The dynamical instability of ferromagnetic condensates that leads to spontaneous formation of spin domains
has been investigated theoretically~\cite{Robins,Zhang_PRL_2005,Ueda_Defect} and observed in experiment~\cite{Sadler_Nat_2006}. An analogous phenomenon has been predicted recently for polar condensates in presence of magnetic field~\cite{Matuszewski_PRA_2008}. Here we correct the results of Ref.~\cite{Matuszewski_PRA_2008}, by noting that
the $\rho_0=1$ state is stable in ferromagnetic condensates for $\delta E > 2|c_2| n$, and the 2C ($\rho_0=0$) state is stable in polar BECs if $\delta E < m^2 /2$. Both states become the ground states for these values of parameters. By investigating stability in various ranges of parameters, we are able to formulate a phenomenological law governing the dynamical stability of condensates:
\begin{enumerate}
\item The only stable state for both polar and ferromagnetic BECs in finite magnetic field is the ground state, as shown in Table~\ref{TGS}.
\item In zero magnetic field, the same is true for ferromagnetic condensates;
However, all stationary states of polar condensates are dynamically stable
in zero magnetic field \cite{Robins,Zhang_PRL_2005,Matuszewski_PRA_2008}.
\end{enumerate}
The reason for the stability of polar condensates in vanishing magnetic field case is not yet clear.
We note that the polar condensates in weak magnetic field may also be effectively stable on a finite time scale.
As shown in Ref.~\cite{Matuszewski_PRA_2008}, in this latter case the instability growth rate of unstable modes is proportional to the fourth power of the magnetic field strength. The time required for the development of instability may be much longer than the condensate lifetime~\cite{Ho_PRL_1998}.

\section{Conclusions} 
\label{Sec_conclusions}

We have studied the ground state of a spin-1 BEC in the presence of a homogenous magnetic field with and without an external trapping potential.  We have found that without a trapping potential the translational symmetry can be spontaneously broken in polar BEC, with the formation of spin domains in the ground state.  We have determined the ground-state phase diagram in the space of magnetization versus magnetic field divided by density, and demonstrated the different phases, each characterized by the type of nonvanishing components.  We have found good agreement between the numerical calculation of the phase diagram  and the analytical predictions based on the homogenous states.  We have shown that these results may be used to understand the ground state structure in the presence of a trapping potential by mapping the locally varying density in the trap to the homogenous state.  We have found that, depending on the magnetic field, the antiferromagnetic BEC ground state in the trap displays pronounced spin domains for a range of possible experimental conditions. 
Finally, we have discussed the relationship between the phenomenon of phase
separation and the dynamical instability leading to the formation of dynamic
spin textures.

\acknowledgments

This work was supported by the Australian Research Council through
the ARC Discovery Project and Center of Excellence for Quantum-Atom Optics.

\clearpage

\end{document}